\documentclass[twocolumn,pra]{revtex4}%
\usepackage{amsmath}
\usepackage{amssymb}
\usepackage{stmaryrd}
\usepackage{marvosym}
\usepackage{graphicx}
\usepackage{url}
\usepackage{subfigure}
\usepackage{fancyhdr}
\usepackage{amsfonts}%
\DeclareMathAlphabet{\mathwee}{OT1}{cmss}{m}{sl}

\begin{document}
\title{Near-field cavity optomechanics with nanomechanical oscillators}
\author{G.\ Anetsberger$^{1}$, O.\ Arcizet$^{1}$, Q.\ P.\ Unterreithmeier$^{2}$, R.\ Rivi\`ere$^{1}$, A.\ Schliesser$^{1}$,
E.\ M.\ Weig$^{2}$, J.\ P.\ Kotthaus$^{2}$, T.\ J.\ Kippenberg$^{1,3}$}
\email{tobias.kippenberg@epfl.ch}
\affiliation{$^{1}$ Max-Planck-Institut f{\"u}r Quantenoptik, Hans-Kopfermann-Str. 1, 85748 Garching, Germany\\
$^{2}$ Fakult{\"a}t f{\"u}r Physik and Center for NanoScience (CeNS), Ludwig-Maximilians-Universit{\"a}t (LMU),
Geschwister-Scholl-Platz 1, 80539 M{\"unchen}, Germany\\
$^{3}$ Ecole Polytechnique F$\acute{e}$d$\acute{e}$rale de Lausanne, EPFL, 1015 Lausanne, Switzerland.}

\small

\begin{abstract}
\bf{
\noindent
Cavity-enhanced radiation pressure coupling between optical and mechanical degrees of freedom allows quantum-limited position measurements and gives rise to dynamical backaction enabling amplification and cooling of mechanical motion. Here we demonstrate purely dispersive coupling of high Q nanomechanical oscillators to an ultra-high finesse optical microresonator via its evanescent field, extending cavity optomechanics to nanomechanical oscillators. Dynamical backaction mediated by the optical dipole force is observed, leading to laser-like coherent nanomechanical oscillations solely due to radiation pressure. Moreover, sub-fm/Hz$^{1/2}$ displacement sensitivity is achieved, with a measurement imprecision equal to the standard quantum limit (SQL), which coincides with the nanomechanical oscillator's zero-point fluctuations. The achievement of an imprecision at the SQL and radiation-pressure dynamical backaction for nanomechanical oscillators may have implications not only for detecting quantum phenomena in mechanical systems, but also for a variety of other precision experiments. Owing to the flexibility of the near-field coupling approach, it can be readily extended to a diverse set of nanomechanical oscillators and particularly provides a route to experiments where radiation pressure quantum backaction dominates at room temperature, enabling ponderomotive squeezing or QND measurements.}

\end{abstract}

\maketitle

Nanomechanical oscillators \cite{CraigheadScience00,Ekinci05} possess wide-ranging applications in both fundamental and applied sciences.
Due to their small mass they are ideal candidates for probing quantum limits of mechanical motion in an experimental setting. Moreover, they are the basis of various precision measurements \cite{Jensen08,Cleland98} and integral part of atomic and magnetic force microscopy \cite{Rugar04} that are pivotal tools for solid state physics and material science. Significant attention has been devoted to developing sensitive readout techniques for nanomechanical motion over the past decade. A natural scale for comparing the performance achieved with systems of different size and mass is given by the variance of the mechanical oscillators' zero-point motion $\langle x\left(t\right)^2\rangle _\mathrm{zp} = \hbar/(2 m \Omega_\mathrm{m})$ ($\hbar$: reduced Planck constant; $m$, $\Omega_\mathrm{m}/2\pi$, $Q$: mass, resonance frequency, quality factor of the oscillator). In Fourier space the zero-point motion can be described by the single-sided (double-sided) spectral density $S_{xx}\left[\Omega\right]$, which at the mechanical oscillator's resonance evaluates to $S_{xx}\left[\Omega_\mathrm{m}\right]=2 \hbar Q/m\Omega_\mathrm{m}^2$ ($S_{xx}\left[\Omega_\mathrm{m}\right]=\hbar Q/m\Omega_\mathrm{m}^2$) and coincides with the standard quantum limit \cite{BraginskyB92, Tittonen99,Caves81,Clerk08} (SQL) of continuous position measurement. So far, the most sensitive transducers for nanomechanical motion have been based on electron flow using a single electron transistor \cite{LaHaye04} (SET) or atomic point contact \cite{Flowers07} (APC) coupled to a nanomechanical string in cryogenic environment and have achieved a position imprecision of order $10^{-15} \,\mathrm{m/Hz^{1/2}}$. An imprecision at the level of the SQL, however, has not been achieved yet.
In contrast, parametric motion transducers based on photons in a cavity-which are the basis for laser gravitational wave interferometers-provide quantum-limited measurement imprecision exceeding $10^{-18} \,\mathrm{m/Hz^{1/2}}$ \cite{ArcizetPRL06,SchliesserNJP08} being at or even below \cite{SchliesserNP08} the zero-point fluctuations of the respective mechanical oscillator but are typically orders of magnitude less sensitive when applied to less massive, nanomechanical oscillators due to the optical diffraction limit \cite{Unter09}. However, cavity-optomechanical coupling of mechanical oscillators allows exploiting radiation pressure dynamical backaction \cite{BraginskyB77,Kippenberg08} that is associated with the momentum transfer from the photons involved in the measurement to the probed object and provides a mechanism for cooling \cite{ArcizetNature06,GiganNature06,SchliesserPRL06} or coherent amplification \cite{KippenbergPRL05} of mechanical motion. As such, an ideal platform would combine the quantum-limited detection and control afforded by cavity-optomechanical coupling with nanoscale mechanical systems which, owing to their small masses provide large zero point motion and high force sensitivity. Such an approach may thereby have promising implications for probing quantum phenomena of mechanical systems \cite{SchwabPT05} and equally in precision experiments such as mass spectroscopy \cite{Jensen08}, charge sensing \cite{Cleland98} and single spin detection \cite{Rugar04} that are based on ultra-sensitive nanomechanical oscillators.

\begin{figure}[h!]
\includegraphics[width=3.2in]{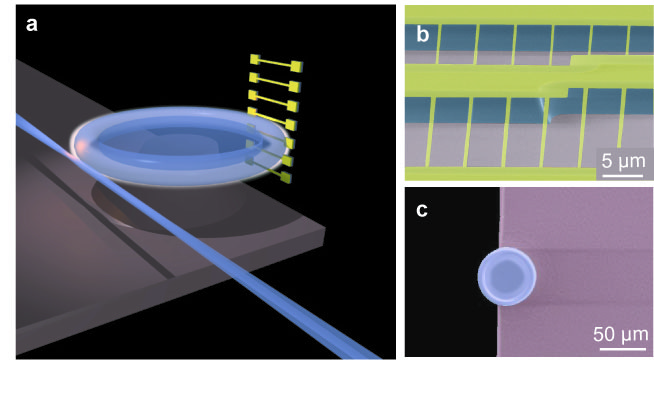}
\caption{\footnotesize{ \textbf{Evanescently coupling a nanomechanical oscillator to an optical microresonator.} \textbf{a,} Schematic of the experiment showing a tapered fibre interfaces optical microresonator dispersively coupled to an array of nanomechanical oscillators. \textbf{b,} Scanning electron micrograph (false colour) of an array of doubly clamped SiN nanostring oscillators with dimensions $110\,\mathrm{nm}\times(300$-$500)\,\mathrm{nm}\times(15$-$40)\,\mathrm{\mu m}$. \textbf{c,} Scanning electron micrograph (false colour) of a toroid silica microcavity acting as optomechanical near-field sensor.}}%
\label{f:1}%
\end{figure}
Efficient optomechanical interaction with nanomechanical oscillators however requires avoiding introducing losses to the high-finesse optical cavity by the nanomechanical object, while maintaining large optomechanical coupling and mitigating thermal effects. Here we demonstrate this combination, via evanescently coupling high-Q nanomechanical oscillators to the tightly confined optical field of ultra-high finesse toroidal silica microresonators. Purely dispersive radiation pressure coupling to the nanomechanical strings is observed and allows sub-fm/Hz$^{1/2}$ displacement imprecision (at room temperature) which equals the standard quantum limit, thus exceeding the performance of the best nanomechanical motion detectors \cite{LaHaye04,Flowers07, Etaki08, Poggio08,Eichenfield09}. In contrast to the recently developed optomechanical zipper cavities \cite{Eichenfield09} which also operate at the nanoscale, the reported near-field approach moreover decouples optical and mechanical degrees of freedom and thus provides a versatile platform to which diverse nanoscale oscillators, such as nanowires \cite{Cui01}, graphene sheets \cite{Bunch07}, or carbon nanotubes can be tunably coupled, extending cavity optomechanics \cite{Kippenberg08} into the realm of nanomechanical oscillators. In particular, it enables simultaneously high mechanical and optical Q giving access to the resolved sideband regime \cite{SchliesserNP08,Wilson-Rae07,Marquardt07}. By detuned excitation, dynamical backaction mediated by the optical dipole force is demonstrated which leads to radiation pressure induced coherent oscillations of the nanomechanical oscillator while thermal effects are negligible. Equally important, the combination of picogram and high quality factor nanostrings \cite{Verbridge08} with ultra-high optical finesse microresonators provides a route to the remarkable regime where radiation pressure quantum backaction is the dominant force noise on the mechanical oscillator even at room temperature and might thus enable quantum optomechanical experiments such as ponderomotive squeezing \cite{Fabre94}, quantum non-demolition (QND) measurements of photons \cite{Heidmann97, Verlot09} or optomechanical entanglement \cite{Vitali07} at ambient temperature.

Fig. \ref{f:1} a shows a schematic of the experimental setup. We employ an array of nanomechanical oscillators in the form of high-Q, tensile stressed and doubly clamped SiN strings \cite{Verbridge08,Unter09} such as depicted in Fig. \ref{f:1} b. The strings have typical dimensions of 

\begin{figure}[b!]
\includegraphics[width=3.5in]{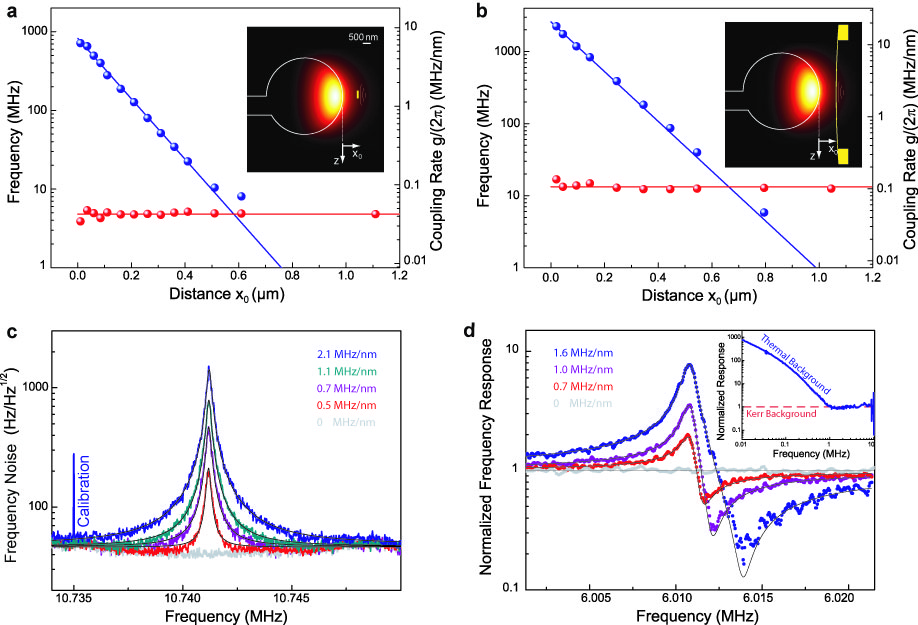}
\caption{\footnotesize{ \textbf{Characterization of the optomechanical coupling.} \textbf{a,} and \textbf{b,} show the dependence of a 58-$\mathrm{\mu}$m-diameter optical microcavity's linewidth (red) and negative optical frequency shift (blue) as a function of the distance $x_\mathrm{0}$ to nanomechanical oscillators in form of a doubly clamped SiN string ($\mathrm{110\,nm\times 800\,nm\times25\,\mu m}$, a) and a 2-D $\mathrm{Si_3 N_4}$ sheet ($\mathrm{30\,nm\times40\,\mu m \times50\, \mu m}$, b). The data reveals in both cases purely dispersive coupling without introducing a measurable degradation of the microcavities' optical decay rate. The right axes show the static optomechanical coupling rates $g\left(x_\mathrm{0}\right)=d\omega_\mathrm{0}\left(x_\mathrm{0}\right)/dx_\mathrm{0}$, as given by the negative derivative of the fitted frequency shift data (blue only). Coupling rates $g/2\pi$ of order $10\,\mathrm{MHz/nm}$ are achieved. \textbf{c,} shows the Brownian noise associated with the nanomechanical oscillator of panel a transduced via the optical cavity for different oscillator positions. Shown as inset are the respective dynamical coupling rates $g/2\pi$ derived from the calibrated frequency noise spectra $S_{\omega\omega}\left[\Omega\right]$ as explained in the text. \textbf{d,} shows the interference of the nanomechanical oscillator's force and the microcavity's Kerr response to a modulated laser field confirming the attractive nature of the dipole force. Moreover, this measurement represents a third, independent method to determine the optomechanical coupling rates (shown as inset: $g/2\pi$, cf. text and SI). Inset: Broadband response of the nanomechanical oscillator-microcavity system, showing thermal cut-off and Kerr-background as well as first low-Q mechanical modes of the microcavity. It is noted that the presence/absence of the nanomechanical oscillator does not alter this broadband, off-resonant response.}}%
\label{f:2}%
\end{figure}

$110\,\mathrm{nm}\times(300$-$800)\,\mathrm{nm}\times(15$-$40)\mathrm{\mu m}$, effective masses of $m_\mathrm{eff} = 0.9$-$5\,\mathrm{ pg}$ and fundamental resonance frequencies  $\Omega_\mathrm{m}/2\pi=6.5$-$16\,\mathrm{ MHz}$ with mechanical quality factors of $Q=10^4$-$10^5$ (cf. Supplementary Information, SI). Following a special fabrication process (cf. SI) indeed allows using the tightly confined optical modes of toroid silica microcavities as near-field probes (cf. Fig \ref{f:1} c) which interact with the nanomechanical oscillator via their evanescent field decaying on a length-scale of $\alpha^{-1}\approx \left(\lambda/2\pi\right)/\sqrt{n^2-1}$ (i.e. approximately $235\,\mathrm{nm}$ for the refractive index of silica $n=1.45$ and a vacuum wavelength of $\lambda\approx 1550 \,\mathrm{nm}$ employed throughout this work).\\

\noindent \textbf{Optomechanical coupling rate}\\
First, we study the strength of the optomechanical coupling of the nanomechanical oscillators to the optical mode of a 58-$\mathrm{\mu m}$-diameter microcavity (exhibiting an unloaded optical linewidth of $4.9 \,\mathrm{MHz}$, i.e. a finesse of $\mathcal{F}=230,000$). The presence of a dielectric oscillator in the evanescent cavity field, at a distance $x_\mathrm{0}$ to the microresonator surface, can in principle give rise to both a reactive and dispersive contribution to the optical cavity response \cite{Favero08}. The former would be characterized by increased cavity losses due to scattering or absorption, i.e. by a position dependent cavity linewidth $\kappa(x_\mathrm{0})/2\pi$. The latter can be described by an optical frequency shift $(\omega_\mathrm{0}(x_\mathrm{0})-\omega_\mathrm{0})/2\pi$ caused by the increased effective refractive index sampled by the evanescent fraction of the mode ($\omega_\mathrm{0}$ denotes the unperturbed cavity frequency with the nanomechanical oscillator being removed). In order to probe this static interaction, we position the nanomechanical strings tangentially to the optical whispering gallery mode (WGM) trajectory and vary their distance (cf. Fig. \ref{f:2} a, inset) to the cavity using piezoelectric positioners. Note that all experiments are performed in this horizontal configuration of the nanostrings as well as in vacuum with a pressure $<10^{-4}\, \mathrm{mbar}$ unless otherwise specified. As depicted in Fig. \ref{f:2} a, the interaction with a nanomechanical string ($\mathrm{110\,nm\times 800\,nm\times25\,\mu m}$) induces an optical frequency shift which exponentially increases as the distance $x_\mathrm{0}$ is decreased and reaches the GHz range. The measured decay length of $110\,\mathrm{ nm}$ is in good quantitative agreement with the theoretically expected value $1/(2\alpha)$. Importantly, we do not measure any degradation of the optical linewidth (cf. Fig. \ref{f:2} a) even for the strongest coupling. Our measurement accuracy of changes in the cavity linewidth $\Delta\kappa/2\pi<0.5\,\mathrm{MHz}$ allows inferring an upper bound of $0.5 \,\mathrm{ppm}$ equivalent optical loss induced by the SiN nanomechanical oscillator. Thus, the optomechanical coupling is purely dispersive  and can therefore formally be described by the dispersive Hamiltonian $\widehat{H}_\mathrm{0}=\hbar \omega_\mathrm{0}(x_\mathrm{0})\hat{a}^{\dagger}\hat{a}$, where $\hat{a}^{\dagger}\hat{a}$ denotes the intracavity photon number. Linearized for small fluctuations $\hat{x}\ll\alpha^{-1}$ around $x_\mathrm{0}$, e.g. for the Brownian motion of the string placed at $x_\mathrm{0}$, the interaction Hamiltonian reads: $\widehat{H}_\mathrm{int}=\widehat{H}_\mathrm{0}+\hbar g(x_\mathrm{0})\hat{x}\,\hat{a}^{\dagger}\hat{a}$ with the optomechanical coupling rate $g(x_\mathrm{0})=d\omega_\mathrm{0}\left(x_\mathrm{0}\right)/dx_\mathrm{0}$. 

Experimentally, the position dependent optomechanical coupling rates of the nanomechanical string to the microcavity can be obtained by taking the derivative of the measured static frequency shifts, i.e. $g(x_\mathrm{0})=\left.d\omega_\mathrm{0}\left(x\right)/dx\right|_{x=x_\mathrm{0}}$. These statically determined coupling rates reach values up to $g/2\pi=10\,\mathrm{ MHz/nm}$ (cf. Fig. \ref{f:2} a). For comparison, also data obtained with a 2-D nanomechanical oscillator in the form of a $30\,\mathrm{nm}\times40\,\mathrm{\mu m}\times50\,\mathrm{\mu m}$ sheet of $\mathrm{Si_3 N_4}$ is shown, which also reveals purely dispersive coupling (cf. Fig. \ref{f:2} b) of up to $g/2\pi= 20\,\mathrm{MHz/nm}$. 
This sizeable coupling is due to the small mode volume $V_\mathrm{cav}$ of toroid microcavities since the optomechanical coupling scales as $g\propto\left(V_\mathrm{nano}/V_\mathrm{cav}\right)\cdot \omega_\mathrm{0}/l$, where $V_\mathrm{nano}$ is the sampled volume of the nanomechanical oscillator and $l$ a characteristic length scale which in our case is given by the field intensity decay length, i.e. $l=1/(2\alpha)\approx110\,\mathrm{nm}$ (cf. SI for analytic expressions of $g$). Yet higher optomechanical coupling rates can be attained in our system by reducing the size of the microcavity and the wavelength of the employed light. Using an integrated photonic crystal device with a larger ratio $V_\mathrm{nano}/V_\mathrm{cav}$, recently allowed obtaining remarkably large coupling rates of ca. $100\,\mathrm{GHz/nm}$ \cite{Eichenfield09}. It however also entailed the disadvantage that it is generally difficult to obtain high optical Q in photonic crystal cavities such that the resolved-sideband regime where the optical linewidth is comparable to or smaller than the mechanical oscillation frequency could not be achieved in ref. 25. The approach presented here, in contrast, separates optical and mechanical degrees of freedom. Since the nanomechanical oscillators do not induce any measurable losses to the ultra-high finesse microresonators it thus particularly allows combining the toroids' high optical Q ($>10^8$) with high mechanical Q and falls naturally into the resolved-sideband regime which enables ground-state cooling \cite{Wilson-Rae07,Marquardt07} or backaction evading measurements \cite{BraginskyB92}.\\

\noindent \textbf{Transduction of nanomechanical motion}\\
The optomechanical coupling rate transduces the motion of the nanomechanical oscillator's eigenmodes (characterized by the displacement spectral density $S_{xx} [\Omega]$) into frequency noise $S_{\omega\omega} [\Omega]$ of the optical cavity mode. Fig. \ref{f:2} c shows the Brownian motion of a nanostring at room temperature ($\mathrm{110\, nm\times800\,nm \times 25\, \mu m}$, $\Omega_\mathrm{m}=10.74\,\mathrm{MHz}$, $Q=53,000$, $m_\mathrm{eff}=3.6\,\mathrm{ pg}$, $x_\mathrm{rms}=16\,\mathrm{pm}$) imprinted into cavity frequency noise, probed by a laser locked to cavity resonance and calibrated using a known external frequency modulation as in prior work \cite{SchliesserNJP08,SchliesserNP08}. The nanostring's room temperature Brownian noise $S_{xx} [\Omega]$ can thus be used to directly determine the optomechanical coupling $g=\sqrt{S_{\omega\omega} /S_{xx}}$ in a second, independent way. We refer to this as a dynamic measurement. Both for the SiN nanostring and the 2-D $\mathrm{Si_3 N_4}$ nanosheet the values obtained for the dynamically measured coupling rates are in good agreement with the statically determined values (cf. SI).
It is important to note that this identity of static and dynamic coupling rates is in agreement with the expectation for optical dipolar interaction which should give rise to frequency independent optomechanical coupling rates $g$. The non-measurably small optical losses ($<0.5\,\mathrm{ppm}$) induced by the nanostrings also indicate that dissipative coupling mediated by thermal effects via light absorption plays an insignificant role. Indeed, differentiating radiation pressure from thermal effects is a challenge that has eluded researchers for centuries. A prominent example is the light mill which can be driven by thermal heating, rather than by radiation pressure. More recently, thermal effects have been shown to play a significant role in micro- and nanomechanical systems \cite{Eichenfield09,Zalalutdinov01,HoehbergerNature04}. It is, however, only the conservative Hamiltonian of radiation pressure that allows phenomena such as ponderomotive squeezing \cite{Fabre94} or quantum non-demolition (QND) measurements of photons \cite{Heidmann97}. Therefore it is central to clearly identify the origin of the optomechanical interaction.\\

\noindent \textbf{Demonstration of radiation pressure interaction }\\
The optomechanical coupling not only gives rise to a differential cavity frequency shift which transduces the nanostring's mechanical motion but also conveys the per-photon force $-\hbar g\left(x_\mathrm{0}\right)$ inevitably acting on the mechanical degree of freedom, as expected for any linear continuous position measurement.
As a proof that indeed the optical dipole force mediates the optomechanical coupling-rather than thermal effects \cite{Eichenfield09,Zalalutdinov01,HoehbergerNature04}-we carry out a pump-probe measurement which probes the force response of the nanomechanical oscillator. A resonant, intensity modulated pump-laser provides the modulated force $\delta F\left[\Omega\right]=-\hbar g \delta N\left[\Omega\right]$ (where $N$ is the intracavity photon number) acting on the nanomechanical oscillator while a second, weak probe-laser measures the response of the cavity resonance frequency. The measured data (cf. Fig. \ref{f:2} d) consists of the nanomechanical oscillator's force response, interfering with the constant background due to the Kerr-nonlinearity of silica (i.e. its intensity dependent refractive index, cf. Fig. \ref{f:2} d and Ref. 20).  Note that the data for different optomechanical coupling rates are scaled to the constant Kerr-background which allows an accurate determination of the magnitude of the nanomechanical oscillator's response.
Two important conclusions can be drawn from this measurement. First, the shape of the interference in Fig. \ref{f:2} d implies that the force experienced by the nanomechanical oscillator is \textit{attractive} (i.e. pointing towards higher intensity) as expected for an optical gradient force. Below its resonance frequency the mechanical oscillator responds in phase with the modulated force. The attractive force then pulls it towards the optical mode which leads to an increased redshift, adding to the in-phase redshift due to the Kerr contribution. Above its resonance frequency where the nanomechanical oscillator responds with a phase lag of 180 degrees, the attractive force leads to destructive interference with the in-phase Kerr response. Second, the ratio of the mechanical to the Kerr nonlinearity response constitutes a relative measure \cite{SchliesserPRL06} and allows deriving the per-photon force acting on the nanomechanical oscillator independent of the optical parameters (cavity linewidth, coupling conditions, input power). The coupling rates $g$ independently measured in this per-photon-force measurement match the values determined by both methods presented earlier (cf. SI for more details). This measurement thus unambiguously demonstrates that the interaction of the nanomechanical oscillator and the optical cavity is mediated by the dispersive, ponderomotive radiation pressure interaction, i.e. the optical dipole force, while no thermal forces are observable.\\

\begin{figure}[h!]
\begin{center}
\includegraphics[width=3.2in]{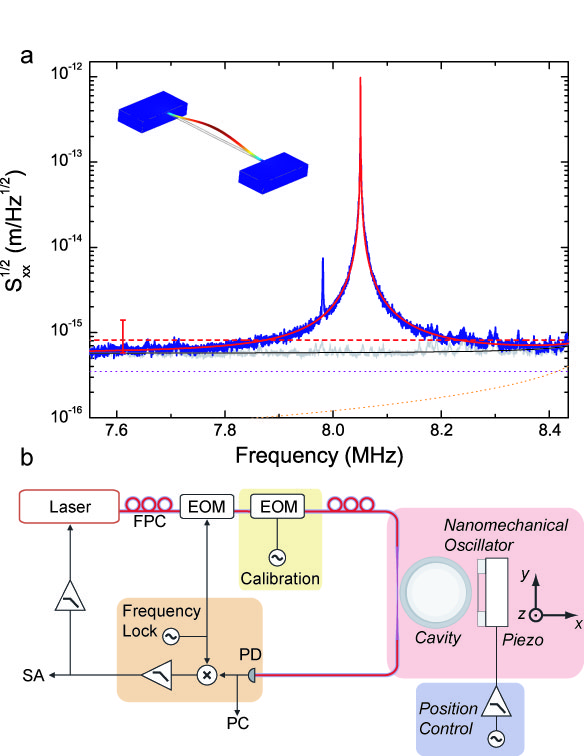}
\end{center}
\caption{\footnotesize{ \textbf{Displacement measurement of a nanomechanical oscillator with an imprecision at the standard quantum limit.} \textbf{a,} Room temperature Brownian noise of a nanomechanical string with a fundamental resonance frequency of $8\, \mathrm{MHz}$ and dimensions $\mathrm{110\,nm\times 800\,nm\times 35\,\mu m}$ ($m_\mathrm{eff}=4.9\, \mathrm{pg}$, $Q=40,000$). For an input power of $65\,\mathrm{ \mu W}$ the displacement imprecision reaches a value of $570\, \mathrm{am/\sqrt{Hz}}$ (grey line), $\times 0.7$ the standard quantum limit, i.e. the oscillator's expected zero point fluctuations of $820\,\mathrm{am/Hz^{1/2}}$ (red dashed line). The large dynamic range across the wide frequency window gives rise to a $1.5\,\mathrm{dB}$ error bar for this value (shown in red). The background is given to 40\% by laser shot-noise (purple dotted line), by thermal noise of the cavity (orange dotted line) and by detector noise (not shown). The second, $20\,\mathrm{dB}$ smaller peak is attributed to a second resonator in the toroid's field of view. Inset: finite element simulation of the string's fundamental mode. \textbf{b,} shows a schematic of the measurement setup used to attain an imprecision at the SQL employing a low-noise fibre laser emitting at $1548\,\mathrm{nm}$ and locked to cavity resonance ($\kappa/2\pi= 50\, \mathrm{MHz}$ for the measurement shown in a). PD: photodiode, PC: personal computer, SA: spectrum analyzer, EOM: electro optic modulator, FPC: fibre coupled polarization controller.}}%
\label{f:3}%
\end{figure}

\noindent \textbf{Measurement of nanomechanical motion with an imprecision at the SQL}\\
Having established its ponderomotive origin, we use the optomechanical coupling to obtain a high-sensitivity readout of nanomechanical motion with an imprecision at the SQL. To this end, we employ a low-noise fiber laser which is quantum-limited in both amplitude and phase at the Fourier frequency of the mechanical oscillators ($>6 \,\mathrm{MHz}$), resonantly locked via a Pound-Drever-Hall technique \cite{ArcizetPRL06}. The experimental setup is depicted in Fig. \ref{f:3} b. Remarkably, using a $\kappa/2 \pi =50\, \mathrm{MHz}$ optical mode, more than $60 \,\mathrm{dB}$ signal to background can be obtained when measuring the Brownian motion $S_{xx}\left[\Omega_\mathrm{m}\right]$ of nanomechanical strings at room temperature. In an ideal measurement the background of the measurement is only given by laser shot-noise, which limits the single-sided displacement sensitivity attainable for an input power $P_\mathrm{in}$ and an impedance matched cavity to \cite{ArcizetPRL06}:
\begin{eqnarray}
\sqrt{S_{xx} \left[\Omega\right]}=\sqrt{\frac{\hbar \omega_\mathrm{0}}{P_\mathrm{in}}} \frac{\kappa/2}{\sqrt{2} g}\sqrt{1 + 4\frac{\Omega^2}{\kappa^2}}\, .
\end{eqnarray}

The best single-sided displacement sensitivity (as determined from the background of the measurement) that was achieved amounts to $S_{xx}^{1/2}= 570\,\mathrm{am/Hz^{1/2}}$, as shown in Fig. \ref{f:3} a, using an 8-MHz doubly clamped nanomechanical SiN string ($110\,\mathrm{nm\times800\,nm\times 35\,\mu m}$, $m_\mathrm{eff}=4.9\,\mathrm{ pg}$, $Q=40,000$), $65\,\mathrm{ \mu W}$ input power and a coupling rate of $g/2\pi= 3.8\, \mathrm{MHz/nm}$.

In order to allow comparing the attained imprecision to values obtained with other nanomechanical motion transducers, we scale it to the nanomechanical oscillator's zero-point fluctuations which for the nanostring of Fig. \ref{f:3} evaluate to $820\,\mathrm{am/Hz^{1/2}}$ (single-sided). Thus, our measurement imprecision amounts to only $\times 0.7$(+0.3/-0.2) the oscillator's zero-point fluctuations, i.e. $\times 0.7$(+0.3/-0.2) the SQL. Interestingly, the condition for a measurement with an imprecision better than the zero-point fluctuations can (for both single- and double-sided spectra) be recast into the condition of a signal to background ratio greater than $\sqrt{2 k_B T/\hbar \Omega_\mathrm{m}}\cong\sqrt{2\bar{n}}$, where $\bar{n}\cong k_B T/\hbar \Omega_\mathrm{m}$ denotes the average phonon occupation number of the mechanical mode ($T$: temperature, $k_B$: Boltzmann constant). Such an imprecision had so far never been achieved, neither with the best transducers of nanomechanical motion based on electronic current flow using SET \cite{LaHaye04}, APC \cite{Flowers07} and superconducting quantum interference device \cite{Etaki08} sensors operating in cryogenic environment nor with integrated photonic crystal systems \cite{Eichenfield09}. Thus, although higher absolute sensitivity has recently been obtained for a $\times10$ more massive oscillator \cite{Eichenfield09}, our approach for the first time allows measuring nanomechanical motion with an imprecision at the SQL. 

While the shot-noise limit extracted from the current measurement is as low as $380 \,\mathrm{am/ Hz^{1/2}}$ (in good agreement with the theoretical expectation of $260 \,\mathrm{am/ Hz^{1/2}}$, cf. SI), our measurement is presently also partially limited by detector noise, which can be eliminated via straightforward technical improvements. Moreover, further improvements are readily feasible. Using smaller microcavities and shorter optical wavelength may allow an increase of $g$ by up to one order of magnitude. Thus, displacement sensitivities at the level of $10^{-17}\,\mathrm{m/Hz^{1/2}}$ may be attainable which would allow measurements with an imprecision far below the SQL. Ultimately, the background afforded by mechanical modes of the cavity \cite{SchliesserNJP08,Anetsberger08} (cf. Fig. \ref{f:3} a) and thermorefractive noise \cite{SchliesserNJP08,Gorodetsky04} of the cavity will limit the sensitivity. These noise sources can, however, in principle be suppressed by cryogenic operation, which toroid microcavities have been shown to be compatible with \cite{Arcizet09}.\\ 

\noindent \textbf{Radiation pressure dynamical backaction}\\
\begin{figure}[b!]
\begin{center}
\includegraphics[width=3in]{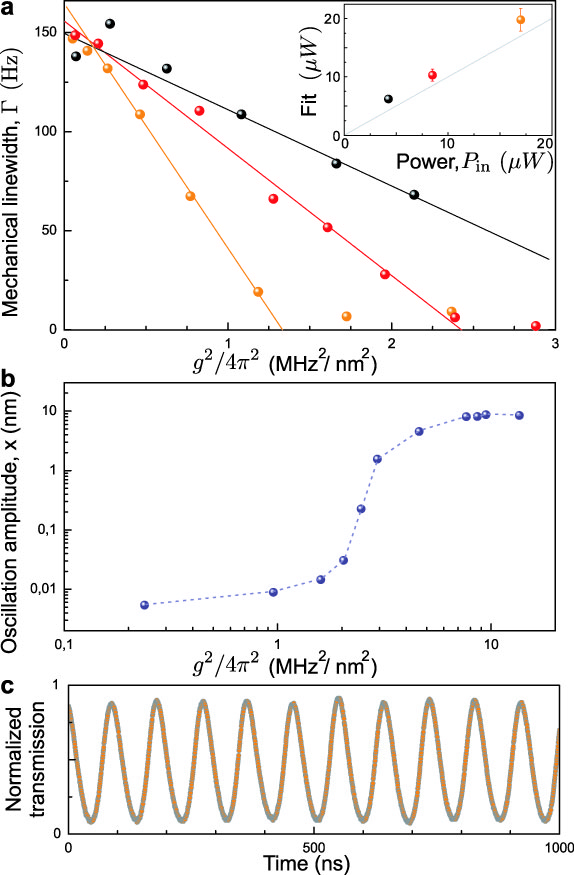}
\end{center}
\caption{\footnotesize{ \textbf{Observation of radiation pressure induced dynamical backaction and coherent oscillations of a nanomechanical oscillator.} \textbf{a,} Mechanical linewidth of a SiN nanostring ($110\,\mathrm{nm}\times 800\,\mathrm{nm}\times 25\,\mathrm{\mu m}$ with a $Q$ of 70,000 at $10.8\,\mathrm{MHz}$ and $m_\mathrm{eff}=3.6\, \mathrm{pg}$) as a function of the optomechanical coupling rate for three different launched powers but fixed blue-detuning $\Delta=+\kappa/2$ ($\kappa/2\pi= 12\, \mathrm{MHz}$). The lines are fits to the dipolar force contribution using the input power as only fit parameter which is in good agreement with the actual input power used (inset). Regions where the linewidth drops to a value close to zero coincide with the onset of regenerative mechanical oscillations. \textbf{b,} shows the oscillation amplitude of the nanomechanical string (derived from a 30-Hz-bandwidth power measurement) as a function of the optomechanical coupling exhibiting threshold and saturation at typical values of $10\,\mathrm{ nm}$. \textbf{c,} shows the transmitted power past the cavity (normalized by the off-resonant transmission) for a nanomechanical string in the regime of the parametric oscillation instability. The coherent mechanical oscillation of the $3.6\, \mathrm{pg}$ string at $10.8\,\mathrm{ MHz}$ causes a close to unity modulation depth of the optical field transmitted by the cavity.}}%
\label{f:4}%
\end{figure}
A second important ramification of the reported cavity-optomechanical system stems from the fact that the nanomechanical oscillators exhibit oscillation frequencies that equal or even exceed the photon decay rate of the optical resonator, enabling access to the regime of dynamical backaction both in the Doppler \cite{ArcizetNature06,GiganNature06,SchliesserPRL06} and resolved-sideband \cite{SchliesserNP08,Wilson-Rae07,Marquardt07} limits. To observe dynamical backaction, the optical microcavity is excited with a positive laser detuning $\Delta=\omega-\omega_\mathrm{0}(x_\mathrm{0})$, which can lead to maser/laser like amplification \cite{Schawlow58} of mechanical motion. Thereby, the mechanical oscillator resumes the role of the photon field in the laser and the cavity, in turn, plays the role of the (phonon) gain medium. As in the case of a laser, the canonical signs of this phenomenon are linewidth narrowing, threshold behaviour and eventual saturation of the oscillation.  All these features are observed with the nanomechanical strings as depicted in Fig. \ref{f:4}. For fixed detuning the backaction gain rate $\Gamma_\mathrm{ba}\propto -g^2 P_\mathrm{in}/m_\mathrm{eff}$ (cf. SI) grows linearly with increased $g^2$. In the experiment the optomechanical coupling $g$ is varied (for fixed detuning $\Delta=\kappa/2$, $\kappa/2\pi=12\,\mathrm{ MHz}$) giving rise to a narrowing of the total mechanical linewidth $\Gamma=\Gamma_\mathrm{m}+\Gamma_\mathrm{ba}$, as shown in Fig. \ref{f:4} a ($\Gamma_\mathrm{m}/2\pi$ denotes the intrinsic mechanical damping rate). The experimentally observed slopes $\partial\Gamma/\partial(g^2)$ are in good quantitative agreement with the theoretical expectation for the dipolar force (cf. inset Fig. \ref{f:4} a). When the backaction rate eventually equals the mechanical damping rate the nanomechanical oscillator experiences net gain which leads to the onset of coherent mechanical oscillations. A clear threshold of the mechanical oscillation amplitude as shown in Fig. \ref{f:4} b is observable, followed by a saturation of the mechanical motion once the frequency shift caused by the mechanical oscillator exceeds the cavity linewidth, leading to gain saturation. Indeed, the large \textit{coherent} oscillations of several nm in amplitude can lead, remarkably and despite the nanoscale nature of the strings, to near-unity modulation depth of the optical cavity transmission as shown in Fig. \ref{f:4} c, when the oscillation amplitude is close to $(\kappa/2)/g$. The resulting RF signal may serve as a photonic clock \cite{hossein-zadeh:023813} and is expected to exhibit a linewidth limited only by thermal noise as in the case of a maser \cite{Schawlow58}. The observation of dynamical backaction amplification (and coherent oscillations) constitutes the first report of dynamical backaction onto a nanomechanical oscillator using radiation pressure (in contrast to thermal effects \cite{Zalalutdinov01}), and in particular using optical gradient or dipole forces. So far, in the field of nanomechanics, dynamical backaction cooling or amplification has only been achieved using microwave fields \cite{Teufel08} which owing to the ca. $\times 10^4$ longer wavelength exhibit lower coupling rates and do not allow access to quantum-limited displacement sensing yet (albeit major progress is currently being made \cite{Castellanos08}).\\

\noindent \textbf{Outlook}\\
The extension of quantum-limited sensitivity with an imprecision at the SQL and dynamical backaction to nanomechanical systems manifests a promising realm for future studies elucidating the quantum nature of optomechanical interaction. Remarkably, we note that combined with state-of-the-art nanomechanical strings \cite{Verbridge08} the ratio of radiation-pressure quantum backaction (the force noise provided by photon shot-noise) and thermal force spectral density can reach unity at room temperature owing to the very small (picogram) mass and ultra-high finesse ($>400,000$) of the optomechanical system (cf. SI). This brings the long sought-after \cite{BraginskyB92, Tittonen99} regime of quantum backaction into reach even at room temperature which would allow ponderomotive squeezing \cite{Fabre94} or QND measurements of the intra-cavity field \cite{Heidmann97,Verlot09}. A further distinguishing feature of the presented approach is that by coupling the nanostrings transversely to the direction of the WGM field, quadratic coupling to the position coordinate of the nanomechanical oscillators enabling QND measurements of mechanical motion \cite{Thompson08,MiaoA09} can be implemented by exploiting the standing wave mode patterns that microresonators can exhibit (cf. SI).

Pertaining to the wider implications, the presented approach allows coupling to virtually any dielectric nanomechanical oscillator. The ability to combine nanoscale mechanical oscillators-which are at the heart of a variety of high-resolution measurement techniques \cite{Jensen08,Cleland98,Rugar04} and proximity (Casimir) force sensors-with quantum-limited displacement sensitivity at the sub-fm/Hz$^{1/2}$ level conceivably offers opportunities for improved performance in these research fields. Particularly interesting may also be the study of graphene sheets \cite{Bunch07}, which have received significant attention in contemporary solid state physics. The possibility to couple several mechanical oscillators to a single optical mode may moreover provide a straightforward way to achieve optically mediated coupling between different mechanical oscillators. Finally, the used microtoroidal platform has already been demonstrated as interface for atomic cavity quantum electrodynamics (cQED) \cite{Aoki06}, enabling potentially the interaction of phonons, photons and atoms or qubits, as recently proposed \cite{Hammerer09,Rabl09}. \\

\noindent \textit{Note added:} After submission of this work a measurement of nanomechanical motion using microwaves with an imprecision at the SQL has been reported \cite{TeufelA09}.


\small
$\hphantom{xxx}$\\
\noindent
\textbf{ Supplementary Information} accompanies the paper.

$\hphantom{xxx}$\\
\noindent
\textbf{Acknowledgements} T.J.K. acknowledges financial support by an Independent Max Planck Junior Research Group of the Max Planck Society, an ERC Starting Grant (SiMP), MINOS and a Marie Curie Excellence Grant as well as the Nanosystems Initiative Munich (NIM). J.P.K. acknowledges financial support by the Deutsche Forschungsgemeinschaft via project Ko 416/18, the German Excellence Initiative via the Nanosystems Initiative Munich (NIM) and LMUexcellent as well as LMUinnovativ. O.A. acknowledges funding from a Marie Curie Intra European Fellowship within FP7 (project QUOM). T.J.K. thanks P. Gruss and the MPQ for continued Max-Planck support. The authors thank M.L. Gorodetsky for valuable discussions.

$\hphantom{xxx}$\\
\noindent
\textbf{ Author contributions} J.P.K. initiated the study and jointly devised the concept with T.J.K.. G.A. and O.A. planned, performed and analyzed the experiments supervised by T.J.K.. Q.P.U. and E.M.W. designed and developed suitable nanomechanical resonators. All authors discussed the results and contributed to the manuscript. R.R. contributed to the development of the experimental apparatus and A.S. assisted with the response measurements.

\widetext
\newpage

\renewcommand{\thefigure}{\textbf{S}\arabic{figure}}
\renewcommand{\theequation}{$\mathrm{S\,} $\arabic{equation}}
\setcounter {figure} {0}
\begin{center}
\large{\textbf{
Supplementary Information - Near-Field Cavity Optomechanics with Nanomechanical Oscillators}}
\end{center}
\vspace{.2in}

\small
\begin{center}
Supplementary Information accompanying the manuscript, pertaining to the
fabrication of the microcavity samples and nanomechanical strings, the optomechanical coupling
coefficient, the optomechanical response measurements, displacement sensitivity and dynamical backaction.
Moreover, prospects for observing quantum backaction at room temperature as well as an elementary demonstration of quadratic coupling are given.
\end{center}
\vspace{.4in}
\section{Fabrication of Optomechanical Near Field Sensors and Nanomechanical Strings}
\subsection{\textit{Optomechanical Near Field Sensors}}
In order to couple toroidal silica microcavities \cite{SIArmani03} to nanomechanical oscillators, the
optical cavity mode has to be readily accessible. To achieve this, optical
microresonators are fabricated at the edge of silicon chips using a suitably
modified microfabrication technique (cf. Fig. S 1). First silica
pads are created on an oxidized silicon wafer using optical lithography and a
hydrofluoric wet-etch. In a second step the chip is cleaved such that the
silica pads lie close to the rim of the chip support. In a subsequent
$\mathrm{XeF}_\mathrm{2}$ dry-etch the full chip with the top side being protected by photoresist 
is monitored by a microscope and at the same
time precisely etched from the side until the edge of the chip lines up with the
edge of the silica pad. After removing the photoresist, the silica pads are
isotropically underetched using a further $\mathrm{XeF}_\mathrm{2}$ dry-etch
which lets the silica pad extend beyond the chip-support. During this phase, an additional silicon chip is placed next to the main sample to ensure homogenous etching. Finally the silica pad is
subject to a $\mathrm{CO}_\mathrm{2}$ laser assisted reflow process creating
the ultra-high finesse microcavity near-field sensor.

\begin{figure}[ptbh]
\centering\includegraphics[width=5.5in]{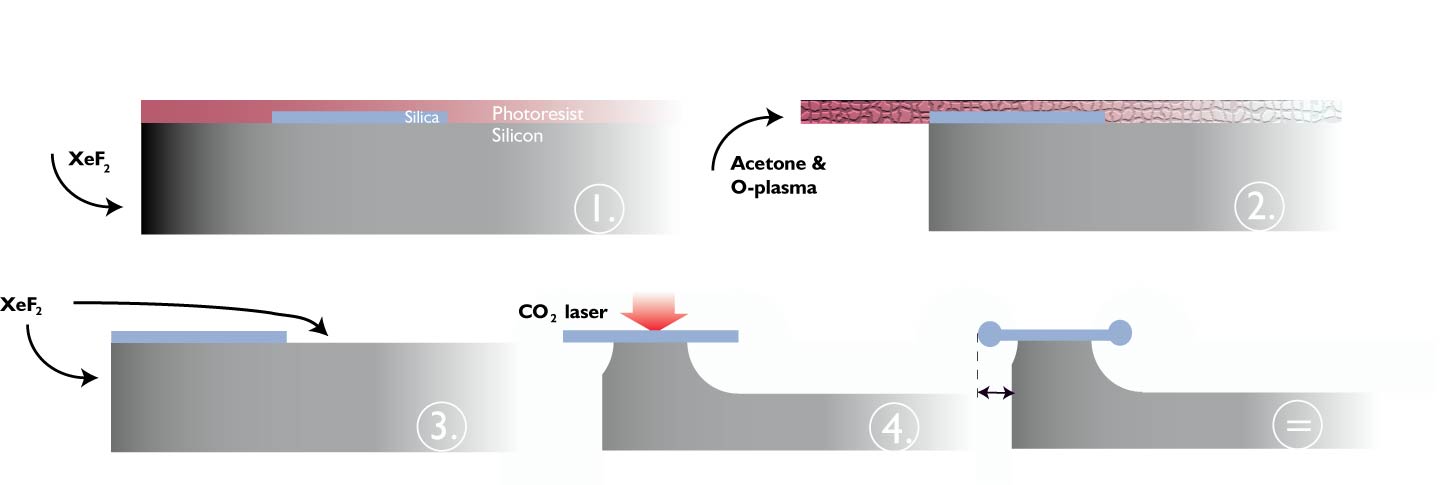}\caption{Sequence for fabricating optomechanical near-field sensors. See text for further information.}%
\label{SIf:1}%
\end{figure}

\subsection{\textit{Nanomechanical Strings}}
The nanomechanical oscillators are fabricated from a high stress silicon nitride film enabling high mechanical quality factors \cite{SIVerbridgeJAP06,SIUnter09}. This $110\,\mathrm{nm}$-thick device layer is commercially grown by LPCVD on top of a $3\,\mathrm{\mu m}$ thick silicon dioxide sacrificial layer hosted by a silicon wafer (available through HSG-IMIT). The sample chip is processed using electron-beam lithography and subsequent reactive ion etching followed by a hydrofluoric wet etch. It contains 20 identical sets of oscillator arrays, each of which holds 144 doubly clamped strings. Their lateral dimensions are varied sampling six lengths from $15\mathrm{\,\mu m}$ to $40\mathrm{\,\mu m}$ and four widths from  $300\mathrm{\,n m}$ to  $1000\mathrm{\,n m}$.

\section{Optomechanical coupling rate of nanomechanical oscillator and optical microcavity}

The Hamiltonian describing the optomechanical coupling \cite{SILaw95} can be written and
linearized in the form:
\begin{equation}
\widehat{H}_{\mathrm{int}}=\hbar\omega_{0}(x_\mathrm{0} + \hat{x})\hat{a}^{\dagger}\hat
{a}\approx H_\mathrm{0}+\hbar g(x_\mathrm{0})\hat{a}^{\dagger}\hat{a}\hat{x}
\end{equation}
where $\hat{x}$ denotes the
mechanical oscillator's position fluctuation around its mean distance to the toroid $x_\mathrm{0}$, $\hat{a}^{\dagger}$ ($\hat{a}$) are creation (annihilation) operators of the optical
mode with position dependent resonance frequency $\omega_{\mathrm{0}}(x_\mathrm{0})$, and the optomechanical coupling rate $g(x_\mathrm{0})$ is
given by:
\begin{equation}
g(x_\mathrm{0})=\frac{\partial\omega_{0}\left(x_\mathrm{0}\right)}{\partial x_\mathrm{0}}\, .
\end{equation}
The linearization is justified when considering the typical amplitudes of the nanomechanical oscillators' Brownian noise of tens of pm at room temperature which is much smaller than the decay length of the evanescent field of typically 200\,nm.
The optical force exerted on the mechanical oscillator by the
intracavity field (power $P$) can be written as
\begin{equation}
\hat{F}=\frac{[\hat{p}_{x},\widehat{H}_{\rm int}]}{i\hbar}=-\hbar g\hat
{a}^{\dagger}\hat{a}=-\hbar g\frac{P}{\hbar\omega_\mathrm{0}}%
\tau_{\mathrm{rt}}\,,
\end{equation}
where $\tau_{\mathrm{rt}}$ denotes the cavity roundtrip time. For a toroid
of radius $R$ and with effective refractive index $n_{\mathrm{eff}}%
=(1-\eta)n+\eta n_{\mathrm{ext}}$ averaged over the mode
profile that entails an evanescent part $\eta$ sampling the refractive index of the surrounding medium, $n_{\mathrm{ext}}$, and the refractive index of silica, $n$, it is given by
$\tau_{\mathrm{rt}}=2\pi Rn_{\mathrm{eff}}/c$ ($c$: speed of light). Note that the force per
photon ($-\hbar g$) is determined solely by the coupling rate $g$. For
comparison, if one studies the coupling to the intrinsic radial breathing mode
of the toroid \cite{SIRokhsariOE05}, then $g=-\omega_{0}/R$ and the force takes
the usual form: $F=2\pi\hbar k\frac{P}{\hbar
\omega_{0}}$, where $k=n_{\mathrm{eff}}\omega_{0}/c$ is the intracavity
effective wave vector.\\

The coupling rate for evanescent optomechanical coupling of a
nanomechanical oscillator to a microresonator can be derived from energy
considerations, calculating the (small) microcavity frequency shift $\Delta\omega_\mathrm{0}(x_\mathrm{0})$ induced by approaching a dielectric oscillator (volume:
$V_{\mathrm{nano}}$) into the evanescent field, compared to the unperturbed cavity resonance with the oscillator being removed, $\omega_\mathrm{0}$:

\begin{equation}
\frac{\Delta\omega_{0}}{\omega_\mathrm{0}}=-\frac{1}{2}\frac{\int_{V_{\mathrm{nano}}}%
{(\epsilon(\vec{{r}})-1)}|\vec{E}(\vec
{r})|^{2}d^{3}\vec{{r}}}{\int_{V}{\epsilon(\vec{{r}%
})\,|\vec{E}(\vec{r})|^{2}\,d^{3}}\vec
{{r}}}\, ,
\end{equation}
where $\epsilon(\vec{r})$ denotes the relative permittivity. 
The microcavity mode volume is important in a variety of studies, such as in cQED \cite{SIKimble1998,SISpillane05} and reads $V_{\rm cav}=\int_{V}
{\epsilon(\vec{r})\,|\vec{E}(\vec
{r})|^{2}\,/}|\vec{E}_\mathrm{max}|^{2}\,d^{3}\vec{r}$, where $\vec{E}_\mathrm{max}$ stands for the
maximum of the electric field. At a distance $x_\mathrm{0}$ outside the cavity's rim, the
evanescent field can be described as $\vec{E}(\vec{x}_\mathrm{0}%
)=\vec{E}_{\mathrm{max}}\xi e^{-\alpha x_\mathrm{0}}$, where $\xi
\vec{E}_{\mathrm{max}}$ represents the field at the cavity/vacuum
interface, that depends on the mode shape and polarization type
(TE/TM) \cite{SILittleJLT99}. The decay constant is approximately given by $\alpha^{-1}%
=\left(\lambda/2\pi\right)/\sqrt{{n}^{2}-1}$. Thus, expression (4) may be approximated by:
\begin{equation}
\frac{\Delta\omega_{0}}{\omega_\mathrm{0}}=-\frac{1}{2}\frac{\tilde{A}_{\mathrm{nano}}}{V_{\rm cav}%
} \frac{1-e^{-2 \alpha t}}{2 \alpha} (n_{\mathrm{nano}}^{2}-1)\xi^{2}e^{-2\alpha x_\mathrm{0}},
\end{equation}
where $\tilde{A}_{\mathrm{nano}}$ denotes the nanomechanical oscillator's area sampled by the
optical field, $t$ its thickness and $n_{\mathrm{nano}}$ is its refractive index. 
If the nanomechanical oscillator is thin enough ($t\ll 1/2 \alpha$), the above expression
can be simplified to:
\begin{equation}
\frac{\Delta\omega_{0}}{\omega_\mathrm{0}}=-\frac{1}{2}\frac{V_{\mathrm{nano}}}{V_{\rm cav}%
}(n_{\mathrm{nano}}^{2}-1)\xi^{2}e^{-2\alpha x_\mathrm{0}}.
\end{equation}
$V_{\mathrm{nano}}$ then denotes the nanomechanical oscillator's volume sampled by the
optical field.
In comparison with cQED \cite{SIKimble1998}, the coupling rate
here scales as the inverse of the microcavity mode volume (instead of its
inverse square root). This is a consequence of the absence of
permanent polarization in our dielectric oscillators. Linearizing eqn. (5) for
small fluctuations $x$ around $x_\mathrm{0}$, one obtains:
\begin{equation}
g=\omega_\mathrm{0}{\alpha}\,\frac{\tilde{A}_{\mathrm{nano}}}{V_\mathrm{cav}}%
\frac{1-e^{-2 \alpha t}}{2 \alpha}(n_{\mathrm{nano}}^{2}-1)\xi^{2}\,e^{-2\alpha x_\mathrm{0}}.
\end{equation}
The typical mode volume for a toroid microcavity can be estimated as
$V_\mathrm{cav}\equiv2\pi R(D_{\mathrm{mode}}/2)^{2}\pi$ where $D_{\mathrm{mode}}$ is
the diameter of the optical mode, whose shape is for simplicity assumed to be circular.

\subsection{\textit{Horizontal configuration for a nanostring oscillator}}

When the nanomechanical oscillator in form of a doubly clamped SiN string is
placed horizontally (i.e. parallel to the propagation of the WGM)--as in the experiments presented in the main manuscript--, one needs to take into account the finite sampling length due to
the toroid's curvature. When the nanostring is positioned tangentially to the
toroid at a distance $x_\mathrm{0}$ from the rim, the field at a volume
element at position $y$ from the center of the nanostring can be expressed as
$E(x_\mathrm{0},y)\propto e^{-\alpha\left(  \sqrt{(R+x_\mathrm{0})^{2}+y^{2}}-R\right)  }$. This
expression yields a mean transverse sampling length of $l_{y}\equiv\sqrt{\pi R/\alpha}$ (with correction terms $\mathcal{O}(x_\mathrm{0}/R,\alpha^{-1}/R)\ll1$). The effective area of the nanomechanical oscillator seen by the light field is
then approximately given by $\tilde{A}_{\mathrm{nano}}^{\mathrm{h}}=w\sqrt{\pi R/\alpha
}$ ($w$: nanomechanical oscillator's width). Thus, 
\begin{equation}
g^{\mathrm{h}}=\omega_\mathrm{0}\alpha\,\frac{2\,w}{\pi^{3/2}
\sqrt{\alpha R}\,D_{\mathrm{mode}}^{2}}%
\,\frac{1-e^{-2 \alpha t}}{2 \alpha}(n_{\mathrm{nano}}^{2}-1)\xi^{2}\,e^{-2\alpha
x_\mathrm{0}}.
\end{equation}
When considering typical parameters, i.e. $R=30\,\mathrm{\mu m}$,
$w=800\,\mathrm{nm}$, $t=110\,\mathrm{nm}$, $\lambda=1.55\,\mathrm{\mu
m}$, $1/\alpha=220\,\mathrm{nm}$, $\xi=0.4$, $n_{\mathrm{nano}}=2.05$ and
$D_{\mathrm{mode}}=3.5\mathrm{\,\mu m}$, one obtains a maximum rate in the
horizontal configuration of $g^{\mathrm{h}}/2\pi \approx 60\,\mathrm{MHz/nm}$, which is within
a factor of 5 of the values obtained experimentally. The discrepancy is ascribed to imperfections in the positioning of the nanomechanical oscillators
with respect to the cavity.\\

\subsection{\textit{Vertical configuration for a nanostring oscillator}}

When the nanomechanical oscillator is placed vertically (i.e. perpendicular to the
propagation of the optical whispering gallery modes, WGM), the oscillator's area
seen by the cavity can be expressed as $\tilde{A}_{\mathrm{nano}}^{\mathrm{v}}=w l_{x}$, where $l_x\equiv\sqrt{\pi r/\alpha}$ ($r$ denotes the toroid's minor radius). This yields the following expression for the coupling rate:
\begin{equation}
g^{\mathrm{v}}=\omega_\mathrm{0}{\alpha}\,\frac{2\,w \sqrt{r/R}}{\pi^{3/2}\sqrt{\alpha R}D_{\mathrm{mode}}^2%
}\frac{1-e^{-2 \alpha t}}{2 \alpha}(n_{\mathrm{nano}}^{2}-1)\xi^{2}\,e^{-2\alpha x_\mathrm{0}}.
\end{equation}
The ratio between the optomechanical coupling rates obtained in the horizontal and vertical
configurations is
\begin{equation}
g^{\mathrm{h}}/g{^{\mathrm{v}}}=\sqrt{R/r}.
\end{equation}
For the typical values specified above and a minor radius of $r=3\,\mathrm{\mu m}$, this expression evaluates to $3$ which is in agreement
with the experimental observation, that slightly higher coupling rates can be achieved in the horizontal configuration.

\subsection{\textit{Nanosheet oscillator}}

For a nanomechanical sheet, the effective area
sampled by the cavity is given by $\tilde{A}_{\mathrm{sheet}}=l_{x} l_{y}$ which for the parameters mentioned above and a sheet thickness of $30\, \mathrm{nm}$ yields a coupling rate of $g/2 \pi \approx 40\, \mathrm{MHz/nm}$ which is only a factor of two from the values obtained experimentally. A more detailed analysis of the actual optical mode profile would allow a more precise comparison to the model.\\

It is interesting to note that the coupling rates exhibit a strong dependence on the optical wavelength, originating in particular from the microcavity mode volume, so that the former could be strongly enhanced when working with smaller wavelengths and cavities.

\section{Effective mass of the nanostring oscillators}

The $n^{\rm th}$ eigenmode of a thin, doubly
clamped string of length $L$, density $\rho$, dominated by its internal tensile
stress $S$ (force per cross-sectional area) oscillates at frequencies approximately given by \cite{SIWeaver90}:
\begin{equation}
\Omega_\mathrm{m}^{\left(n\right)}/2\pi=\frac{n}{2L}\sqrt{\frac{S}{\rho}}\,.
\end{equation}
This allows deducing an internal tensile stress of $0.9\,\rm GPa$ from our measurements.
In this limit the fundamental mode pattern of the strings extending from $y=-L/2$
to $y=L/2$ can be described as
\begin{equation}
u\left(y\right)  =u_{\mathrm{0}}\,\mathrm{cos}\left(  \frac{\pi}%
{L}y\right)  \,
\end{equation}
and fulfills the equipartition theorem in the following sense:
\begin{equation}
\frac{1}{2}m \, \Omega_\mathrm{m}^2 \left\langle u^2\right\rangle=k_B T \, ,
\end{equation}
where $m=\rho\, t w L$ denotes the oscillator's physical mass and $\left\langle u^2\right\rangle=1/L \int_{L}u(y)^2 dy$ is the mean squared displacement amplitude averaged along the string's length (note that for the time and space averaged root mean square amplitude $\bar{u}$ one would obtain $\bar{u}^2=1/2 \left\langle u^2 \right\rangle$).
In general, however, the measured effective coordinate deviates from $\left\langle u^2\right\rangle$ since it is weighted by the geometry of the measurement apparatus which in our case is given by the spatial distribution of the normalized optical mode profile $v_\mathrm{0}(y)$ sampling the nanomechanical string ($\int_{-\infty}^{\infty}{ v_0(y)^2 dy}=1$). This yields an effective measured squared displacement of $\left\langle u\right\rangle^2_{v_\mathrm{0}^2}=\left(\int_{L}u(y) v_\mathrm{0}(y)^2 dy\right)^2$. Thus, the oscillator's mass correspondingly has to be adjusted in order to maintain equipartition for the actually measured coordinate, leading to an individual effective mass $m_{\rm eff}$ for each mechanical mode. The effective mass of the $n$-th mode (with mode pattern $u_n$) can then generally be written as
\begin{equation}
m_{\rm eff}^{\left(n\right)}= m\frac{\left\langle u_n^2\right\rangle}{\left\langle u_n\right\rangle^2_{v_\mathrm{0}^2}} =m\frac{1/L \int_L{u_{n}(y)^2}dy}{\left(\int_L{ u_{n}(y) v_\mathrm{0}(y)^2 dy} \right)^2}\, .
\label{eq:meff}
\end{equation}
Assuming for example a point like measurement, i.e. $v_\mathrm{0}(y)^2=\delta(y)$, the effective mass of the eigenmodes which are symmetric around $y=0$ evaluates to $m/2$. Note that the effective mass of all antisymmetric modes in contrast diverges for any symmetric probing around $y=0$. This is in fact used in practice for optimizing the centering of the nanomechanical strings.\\
Inserting the approximate expression of the toroid field obtained in the previous section $v_0(y)=\left(\frac{\alpha}{\pi R}\right)^{1/4}e^{- \alpha y^2/2R}$ (for the horizontral
configuration as used in the experiments presented in the main manuscript) as well as the fundamental mode pattern of the strings into eqn. (\ref{eq:meff}) one obtains the following expression for the effective mass $m_\mathrm{eff}$ of the fundamental mode:
\begin{equation}
\frac{m_{\rm eff}}{m}= \frac{\pi R}{L \alpha}\frac{ \int_L{\cos^2(\frac{\pi}{L} y)}dy}{\left(\int_L{ \cos(\frac{\pi}{L}y) e^{- \alpha y^2/R} dy} \right)^2}
\end{equation}
or
\begin{equation}
\frac{m_{\rm eff}}{m}= \frac{1}{2} \frac{\beta^{-1}}{\left(\int_{-\pi/2}^{\pi/2}{
\cos(u) e^{-\pi \beta u^2} du} \right)^2}\, ,
\label{eq:meff2}
\end{equation}
where  $\beta^{-1}=\frac{\pi^3 R}{L^2 \alpha}=(\pi l_y/L)^2$ reflects the transverse sampling length $l_y$ compared to the string's length $L$.\\

Taking into account the effective mass of the respective mechanical mode is important to get consistency between static and dynamic optomechanical coupling rates.
The statically determined optomechanical coupling rates ($g=d\omega_\mathrm{0}(x_\mathrm{0})/dx_\mathrm{0}$) are independent of the above considerations, as they are obtained by uniformly displacing the nanomechanical strings. The dynamically determined optomechanical coupling rates ($g=\sqrt{S_\mathrm{\omega\omega}/S_{xx}}$), however, are not, since the transduction of the individual mechanical modes' motion to cavity frequency noise involves the mode pattern of the particular mechanical modes. Thus, the overlap between optical and mechanical modes has to be accounted for in order to recover correct displacement amplitudes. For a point like measurement of e.g. the fundamental mode at the center of the string, the measured coordinate is given by the the peak displacement $u_\mathrm{0}=\sqrt{2 \left\langle u^2\right\rangle}$ which is $\times \sqrt{2}$ larger than the root mean squared displacement $\sqrt{\left\langle u^2\right\rangle}$ averaged along the full length of the string. Thus, the measured mechanical oscillator mode effectively can be described by a $\times 2$ smaller mass, i.e. its effective mass $m_\mathrm{eff}=m/2$ as mentioned above.\\
All our measurements are in fact close to such a point like measurement, i.e. $\beta^{-1}$ is sufficiently small such that the second fraction in eqn. (\ref{eq:meff2}) evaluates to values close to unity yielding effective masses $m_\mathrm{eff}\approx m/2$. The deviations from the point-like value $m_\mathrm{eff}=m/2$ (largest for the shortest strings of $15\, \mathrm{\mu m}$ where a deviation of 30\% is found) are, however, accounted for. Thus employing the correct efffective mass for the single-sided Brownian noise $S_{xx}[\Omega]= 4\, m_\mathrm{eff} \Gamma_\mathrm{m} k_B T \left|\chi_\mathrm{m}[\Omega]\right|^2$
and the susceptibility $\chi_\mathrm{m}[\Omega]$ (
$\chi_\mathrm{m}[\Omega]^{-1}= m_\mathrm{eff}(\Omega_\mathrm{m}^2-\Omega^2-i\,\Gamma_\mathrm{m}\Omega)$)
ensures that static and dynamic coupling rates as well as the ones determined by the force response (cf. next section) match.

\section{Nanomechanical oscillator-cavity response measurement}

\label{s:response}

In contrast to thermal noise measurements, the use of an intensity modulated
optical beam allows extracting information on the phase response of the
nanomechanical oscillator, in particular enabling to identify the direction of
the forces exerted on the oscillator. Here, following previous
work \cite{SISchliesserPRL06}, we describe a pump-probe measurement that proves the attractive nature of the light-force, as expected for an
optical gradient force, and furthermore provides an additional and independent
measurement of the coupling rate $g$.

\begin{figure}[h]
\begin{center}
\scalebox{1.2}{\includegraphics[width=3.5in]{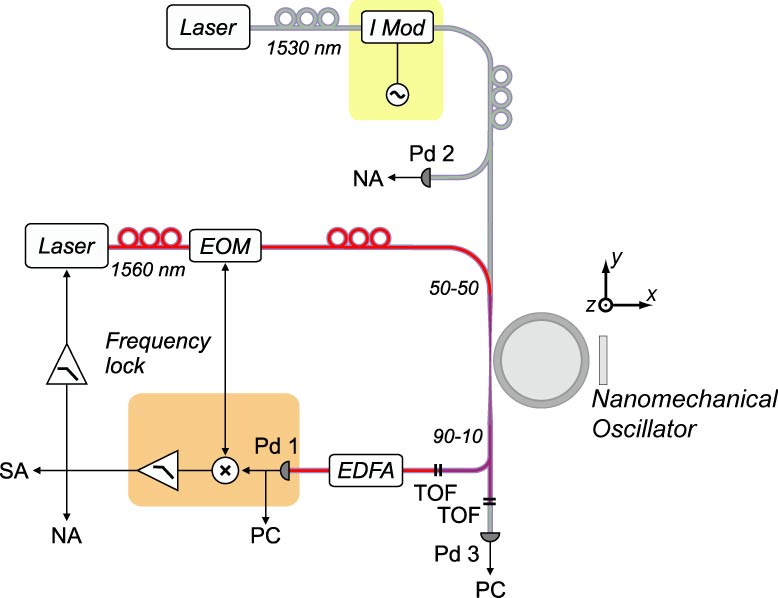} }
\end{center}
\caption{{\small Experimental setup used for the pump-probe measurements. The amplitude modulated pump and the probe beam (both at resonance of two optical modes) are sent to the cavity via a fibre coupled beam combiner. The transmitted signals of both pump and probe beam are accessed individually via a fibre coupled splitter and tunable optical filters (TOF). The probe beam is amplified via an EDFA before detection to allow probing of the cavity-nanomechanical-oscillator system's response to the pump beam (via a Pound-Drever-Hall technique) at low probe power. EDFA: Erbium doped fibre amplifier, I Mod: intensity modulator, P Mod: phase modulator, FPC: fibre polarization controller, Pd: Photodiode, PC: personal computer, SA: spectrum analyzer, NA: network analyzer.}}%
\label{SIf:2}%
\end{figure}

\textit{Experiment ---} Fig. S \ref{SIf:2} shows the experimental setup. We use a fibre-coupled diode laser operating around a wavelength
of $1530\,\mathrm{nm}$ serving as
\emph{pump} laser. It provides a sinusoidally modulated input power $\delta
P_{\mathrm{in}}(t)$, which in turn causes the optical resonances to
periodically shift in frequency via thermal effects due to light absorption,
via light-force induced mechanical displacement of both the toroid and the
nanomechanical oscillator, but also via the Kerr-nonlinearity of
silica. These shifts are read out with a $1560\,\mathrm{nm}$
\emph{probe} laser locked to a microcavity resonance, whose transmitted phase
reflects the induced equivalent cavity resonance frequency shifts. Modulating
the power of the pump laser and demodulating the detected probe error signal with the
same (swept) frequency, we are able to observe and discriminate the different mechanisms contributing to the system's response. 
While electronic modulation and demodulation
are conveniently accomplished with an electronic network analyzer, we use a
fibre-coupled interferometric LiNbO${}_{3}$ amplitude modulator to generate
the modulated pump light. The polarization of both lasers is independently
adjusted prior to their injection in a 50/50 coupler. The light transmitted by the oscillator-cavity system is split using a 90/10 coupler and the separation between the
pump and probe fields is ensured by means of tunable optical filters. The main
port serves for probing as well as locking the probe laser while the weak port
allows maintaining the pump laser at resonance. We verified that residual crosstalk
was more than $10\,\mathrm{dB}$ below the weakest signals detected when
working with sufficiently high Q optical resonances.

\textit{Broadband response ---} First, we measure the broadband response
of the nanomechanical oscillator-cavity system. As shown in Fig. 2 d (inset) of the main manuscript,
it is the sum of several phenomena. At low frequencies, the microcavity response to the intensity
modulation is dominated by thermal mechanisms. The most prominent is the
temperature induced refractive index change that affects the effective length
of the cavity. The thermorefractive contribution strongly reduces with
frequency, and above $1\,\mathrm{MHz}$ one can observe a plateau corresponding
to the silica Kerr effect contribution and a few resonances corresponding to
the first mechanical modes of the toroid (the mechanical modes of the
nanomechanical oscillator are not resolved due to their high mechanical Q). Note that we do
not see a change of any broad-band noise when the cavity is probed with and
without nanomechanical oscillator which allows inferring that the thermal
response of the nano-object is much smaller than the toroid's response.

The Kerr response of silica (i.e. the intensity dependent
refractive index) serves as a reference in our measurements. Under a
modulation $\delta P$ of the intracavity power it causes an equivalent cavity optical resonance frequency change given by:%
\begin{equation}
\delta\omega_{\mathrm{0}}^{\mathrm{Kerr}}[\Omega]= - \omega_{0}\frac{n_{\mathrm{2}}%
}{n_{\mathrm{eff}} A_{\mathrm{mode}}} \, \delta P[\Omega]\, ,
\end{equation}
where $n_{\mathrm{2}}= 3\cdot10^{-16}\,\mathrm{cm}^{2}\,\mathrm{W}^{-1}$ is the Kerr
coefficient of silica and $A_{\mathrm{mode}}$ the optical mode area
($A_\mathrm{mode}\approx\pi(D_{\mathrm{mode}}/2)^{2}$).\newline

\textit{Mechanically resonant displacement ---} The modulated intracavity
power exerts a modulated dipolar force $\delta F[\Omega]$ on the nanomechanical oscillator that
in turn modulates the microcavity optical resonance according to:%

\begin{equation}
\delta\omega_{0}^{\mathrm{nano}}[\Omega]=g_{\mathrm{probe}}\delta
x[\Omega]=g_{\mathrm{probe}}\chi_{\mathrm{m}}[\Omega]\delta F[\Omega]\, ,
\end{equation}
where $g_{\mathrm{probe}}$ is the coupling rate of the probe optical mode and $\chi_{\mathrm{m}}[\Omega]=\frac{1}{m_{\mathrm{eff}}(\Omega_{\mathrm{m}%
}^{2}-\Omega^{2}-i\Omega\Gamma_{\mathrm{m}})}$ and $\delta x[\Omega]$ are the susceptibility
and the oscillation amplitude of the nanomechanical oscillator. Note our definition of $\delta x$,
i.e. the coordinate $\delta x$ increases, when the nanomechanical oscillator moves away
from the microcavity, and thus the optical resonance frequency is
blue-shifted. The modulated optical force applied on the oscillator can be
written as:
\begin{equation}
\delta F[\Omega]=-\hbar g_{\mathrm{pump}}\frac{2\pi Rn_{\mathrm{eff}}}{c}%
\frac{\delta P[\Omega]}{\hbar\omega_{0}} \, .
\end{equation}
We have differentiated between the coupling parameters for the pump ($g_{\mathrm{pump}}$)
and probe ($g_{\mathrm{probe}}$) modes since they exhibit different spatial profiles. Note also that we assumed in this expression that the effective mass in the response measurement is the same as the one defined previously. Both coupling
parameters have been separately measured, exploiting their respective induced
static optical frequency shift. The intracavity power is linked to the injected modulated power
$\delta P_{\mathrm{in}}$ by $\delta P=\frac{\mathcal{F}}{\pi} \delta P_{\mathrm{in}}$ in the case of critical
coupling, where $\mathcal{F}=c/(n_\mathrm{eff} R \kappa)$ denotes the optical finesse (the linewidth $\kappa/2\pi$ of the pump
mode is $120\, \mathrm{MHz}$, such that the cavity cut-off can be safely neglected for the
modulation frequencies below $15\,\mathrm{MHz}$ considered here).

The total response of the system is the coherent sum of the various mechanisms
contributing. Since we are only interested here in the local response in the
vicinity of the nanomechanical oscillator's resonance, we only consider
Kerr and mechanical responses:
\begin{equation}
\frac{\delta\omega_{0}^{\mathrm{tot}}}{\delta\omega_{0}^{\mathrm{Kerr}}%
}[\Omega]= 1+\frac{ g_{\mathrm{probe}}g_{\mathrm{pump}}}{\omega_{0}^{2}}%
\frac{2\pi R n_{\mathrm{eff}}^{2} A_{\mathrm{mode}}}{c n_{2} } \chi
_{\mathrm{m}}[\Omega] \, .
\end{equation}

The frequency response data, normalized to the Kerr background, were fitted using the model
\begin{equation}
H[\Omega]=\left\vert 1+ \frac{a_{1}}{\Omega_{\rm m}^{2}-\Omega^{2}-i\Omega\Gamma_\mathrm{m}
}\right\vert ,
\end{equation}
that represents the coherent sum of a damped driven harmonic mechanical
oscillator and a unity (Kerr) background response. Note that this allows for
interference to occur between the responses of the mechanical modes and the
instantaneous Kerr response, since the former gets out of phase with the latter when the
driving modulation is swept over the resonance frequency (at resonance the
mechanical oscillator response is $\pi/2$ out of phase with the driving
force) giving rise to a dispersive shape in the global response
curve. As shown in Fig. 2 of the main manuscript, this effect is indeed
observed with an excellent agreement between the data and the fit model. No dependence on the modulation amplitude could be observed. 

\begin{figure}[t!]
\centering\includegraphics[width=3.5in]{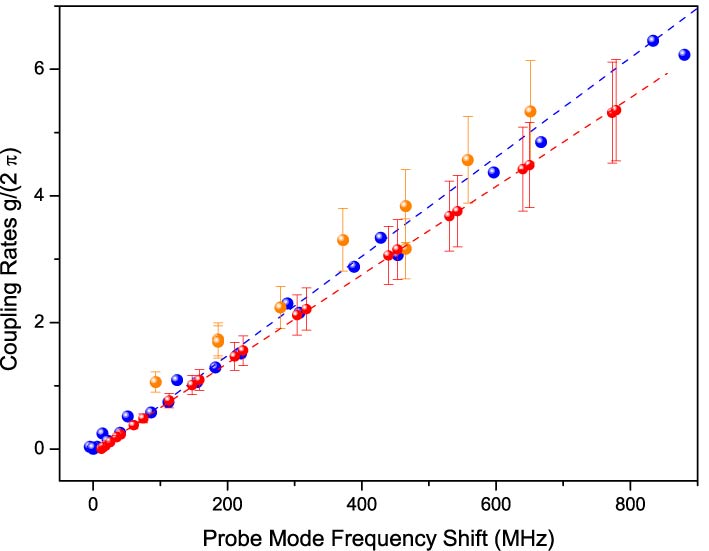}\caption{Shown are the
coupling rates $g_\mathrm{pump}$ (red) and $g_\mathrm{probe}$ (blue) which are evaluated from static measurements as well as the
effective coupling rate $g_\mathrm{eff}$ determined by the force-response measurement (orange). Error bars represent the uncertainties in scaling
the different data sets to a common base. To within our measurement accuracy,
we confirm the magnitude expected for the dipolar force which demonstrates
independently that the force on the nanomechanical oscillator is dominated by the optical
dipole force.}%
\label{SIf:3}%
\end{figure}

The frequency dependence of the mechanical susceptibility, i.e. the fact that
it responds in phase (out of phase) at frequencies lower (higher) than its
eigenfrequency, allows accessing the sign of the measured force. Importantly, the fact that the interference dip is present at frequencies higher
than the mechanical resonance frequency confirms the
\textsc{attractive nature of the optical force}, as expected for a dipolar
force, oriented towards the region of higher optical intensity.

The results from the fits for different oscillator positions (the coefficients
$a_{1}$) allow extracting the effective force-response coupling rate $g_\mathrm{eff}=\sqrt{g_\mathrm{pump}\, g_\mathrm{probe}}$. As shown in Fig. S 3, good agreement at a level of
$10 \%$, within the experimental errorbars, is found between the static and the force-response based determination of the coupling rate $g_\mathrm{eff}$, underpinning the fact that the force acting on
the nanomechanical oscillator is dominated by the optical dipole force.

\section{Displacement Sensitivity}
The quantum nature of light limits the (double-sided) displacement sensitivity to a
level of
\begin{equation}
\sqrt{S_{xx}^{\rm shot}}[\Omega]= \frac{\kappa}{ 4 g}\frac{1}{\sqrt{P_{\rm
in}/\hbar\omega_0}}\sqrt{1+\left(\frac{2\Omega}{\kappa}\right)^2}
\label{equ-Sshot}
\end{equation}
in case of direct phase quadrature read out (homodyne detection),
impedance matching for the cavity and ideal photodetectors. For the parameters $\kappa/2\pi=50\,\mathrm{MHz}$, $g/2 \pi=3.8\,\rm MHz/nm$, $P_{\rm in}= 65\,\rm\mu W$, $\Omega/2 \pi=8\,\rm MHz$, $\lambda=1550\,\mathrm{nm}$ one can achieve a displacement sensitivity of $1.5\times 10^{-16}\,\rm m/\sqrt{Hz}$. In case of
a single-sided Pound-Drever-Hall detection technique, this value is increased by a factor of 1.73.

\section{Dynamical Backaction}

When working with a laser detuned by $\Delta$ with respect to the cavity resonance, the radiation pressure force experienced by the nanomechanical oscillator can become viscous leading to cooling or amplification of motion. In case of critical coupling, the dynamical backaction rate can be written as \cite{SISchliesserPRL06}
\begin{equation}
\Gamma_{\rm ba}=\left(\frac{g x_\mathrm{zp}}{\kappa}\right)^2 \frac{P_{\rm in} }{\hbar\omega_0}
\frac{8}{1+4\Delta^2/\kappa^2}\left(
\frac{1}{1+4\left(\Delta+\Omega_{\rm m}\right)^2/\kappa^2}
-\frac{1}{1+4\left(\Delta-\Omega_{\rm m}\right)^2/\kappa^2}
\right)\,,
\end{equation}
where $x_\mathrm{zp}=\sqrt{\hbar/(2 m_\mathrm{eff} \Omega_\mathrm{m})}$ denotes the nanomechanical oscillator's zero point motion.
For a laser blue detuned to half the side of the fringe of the optical resonance, corresponding to $\Delta=+\kappa/2$, the previous expression can be simplified to:
\begin{equation}
\Gamma_{\rm ba}=- \left(\frac{g x_\mathrm{zp}}{\kappa}\right)^2 \frac{P_{\rm in} }{\hbar \omega_0}
\frac{8 \Omega_\mathrm{m}/\kappa}{1+4\Omega_{\rm m}^4/\kappa^4}\,.
\end{equation}
The data represented in Fig. 4 a of the main manuscript, are fitted with the previous expression for $\Delta=+\kappa/2$. The only free fit parameter is the input power $P_{\rm in}$. When compared to the experimental values, we find good agreement within the error bars of the measurement (see inset in Fig. 4 a of the main manuscript), proving that the backaction observed is dominated by the optical dipole force.\\

The threshold for the onset of the optomechanical parametric instability is reached when the injected power is sufficient to ensure $\Gamma_{\rm ba}=-\Gamma_{\rm m}$:

\begin{equation}
P_{\rm in}^{\rm thres}=\frac{\hbar \omega_\mathrm{0}}{4}\, \frac{m_{\rm eff} \Gamma_{\rm m} \Omega_\mathrm{m}}{\hbar}\frac{\kappa^2}{g^2}\frac{\kappa}{\Omega_\mathrm{m}} \left(1+4\Omega_{\rm m}^4/\kappa^4\right)\, .
\end{equation}

\section{Quantum Backaction on a Nanomechanical String}

The nanomechanical oscillator is both subject to the stochastic thermal Langevin force as well as
to the quantum fluctuations of the optical dipole force induced by the shot-noise of the optical input field, i.e. a quantum backaction force \cite{SIBraginskyB92}. The (double-sided) thermal Langevin force spectral density is given by the
expression:
\begin{equation}
S_{FF}^{\mathrm{th}}[\Omega]=2 m_\mathrm{eff}\, \Gamma_\mathrm{m}\, k_{B}T\, .
\end{equation}

The quantum nature of the intracavity field is responsible for a fluctuating quantum backaction force given by $\delta F[\Omega]=-\hbar g \tau_\mathrm{rt} \delta I[\rm \Omega]$ where $\delta I$ represents the quantum fluctuations of the intracavity photon flux. The noise properties of the latter can be linked to the input flux quantum noise spectrum by $S_I[\Omega]=\frac{P_{\rm in}}{\hbar\omega_0} \frac{\mathcal{F}^2}{\pi^2} \frac{2}{1+4 \Omega^2/\kappa^2}$, in case of critical coupling and for a quantum noise limited input light source.
Using the cavity finesse $\mathcal{F}=c/(n_\mathrm{eff} R \kappa)$, the (double-sided) backaction force spectral density can then be written as:
\begin{equation}
S_{FF}^{\rm qba}[\Omega]= 8 \frac{\left(\hbar g\right)^2}{\kappa^2}  \frac{P_{\rm in}}{\hbar\omega_0} \frac{1}{1+ 4 \Omega^2/\kappa^2}\, .
\label{equ-Sqba}
\end{equation}

Note that the spectral densities (\ref{equ-Sshot}) and (\ref{equ-Sqba}) fulfill the Heisenberg inequality in the form
\begin{equation}
S_{xx}^{\rm shot}[\Omega]\cdot S_{FF}^{\rm ba}[\Omega]=2 \frac{\hbar^2}{4},
\end{equation}
as expected in case of critical coupling.

A key quantum optomechanical experiment would lie in observing the quantum backaction of the measurement laser onto the mechanical oscillator. This would allow observing quantum mechanical correlations between a meter and probe beam and would thus constitute a QND
measurement scheme for optical fields \cite{SIBraginsky80,SIHeidmann97}. For such a measurement, an
obvious prerequisite is to enter a regime where the quantum backaction becomes
comparable to--or even larger than--the thermal Langevin force spectral density,
i.e. where the ratio
\begin{equation}
\frac{S_{FF}^{\mathrm{qba}}[\Omega]}{S_{FF}^{\mathrm{th}}[\Omega]} =   \frac{\hbar}{m_\mathrm{eff} \Gamma_\mathrm{m}\Omega} \left(\frac{ g}{\kappa}\right)^2 \frac{\hbar \Omega}{k_B T}
\frac{P_{\rm in}}{\hbar\omega_0} \frac{4}{1+ 4 \Omega^2/\kappa^2}
\end{equation}

is comparable to unity. In our present experiment this ratio is still small, on the order of $10^{-3}$.
Quite remarkably, however, for an ultra-high Q strained and doubly clamped $\mathrm{Si}%
\mathrm{N}$ string (as those demonstrated by Verbridge et al. in
Cornell with a $Q$ of $1,000,000$ at a frequency of $1\,\mathrm{MHz}$ and
effective mass of $15\,\mathrm{pg}$ \cite{SIVerbridge08}) dispersively coupled
to a microresonator, a promising outlook can be given. To illustrate this point, eqn. (25) is rewritten as
\begin{equation}
\frac{S_{FF}^{\mathrm{qba}}}{S_{FF}^{\mathrm{th}}}[\Omega_\mathrm{m}] \approx   \left(\frac{g/2\pi}{20\, \mathrm{MHz/nm}}\right)^2 \left(\frac{4\, \mathrm{MHz}}{\kappa/2\pi}\right)^2 \left(\frac{15\, \mathrm{pg}}{m_\mathrm{eff}}\right)
\left(\frac{Q}{10^6}\right)  \left(\frac{1\, \mathrm{MHz}}{\Omega_\mathrm{m}/2\pi}\right) \left(\frac{P}{100\, \mathrm{\mu W}}\right)\left(\frac{\lambda}{780\, \mathrm{nm}}\right)\left(\frac{300 \,\mathrm{K}}{T}\right)\,.
\end{equation}

Thus, assuming a decay rate of $\kappa/2\pi=4\,\mathrm{MHz}$ at a wavelength $\lambda=780\,\mathrm{nm}$ and an input power of $P=100\,\mu\mathrm{W}$, this ratio reaches unity for a
coupling rate of $g/2\pi=20\,\mathrm{MHz}/\,\mathrm{nm}$ at \emph{room
temperature}. Such a coupling rate which is a factor 2-3 above the currently
shown rates should in principle be achievable for shorter cavity radius and wavelength.

\section{Quadratic Coupling}
\begin{figure}[t!]
\begin{center}
\scalebox{1.2}{\includegraphics[width=4in]{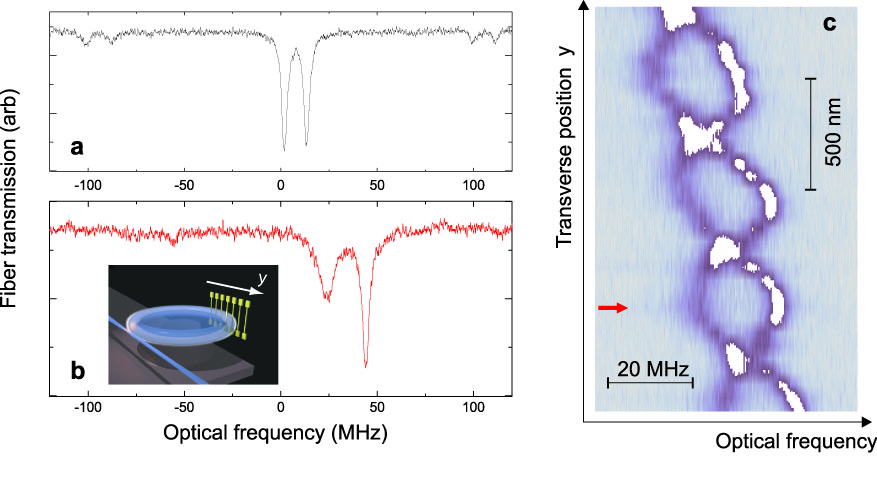} }
\end{center}
\par
\caption{{\small Quadratic coupling of the nanostring's in-plane motion to a split optical microresonator mode when positioning the nanostrings perpendicular to the WGM propagation direction. See text for further information.}}%
\label{SIf:4}
\end{figure}
Optical WGM resonators can enter a regime where the doubly
degenerate clockwise (CW) and counterclockwise (CCW) optical modes form
doublets due to intrinsic scattering centers giving rise to coupled CW
and CCW modes \cite{SIWeiss95,SIKippenbergOL02}. This leads to the
formation of standing wave pairs within the resonator corresponding to the respective non-degenerate new eigenmodes (cf. Fig. S 4a). In the vertical coupling configuration (where the strings are positioned perpendicular to the propagation of the WGM, cf. inset Fig. S 4b) the nanomechanical string can be positioned at a node or antinode of the corresponding standing wave and in this manner, the coupling
of the string's in-plane motion to the optical fields can be rendered quadratic.
Fig. S 4c shows  a color-coded image of the cavity transmission as a function of the lateral position $y$ of a nanomechanical string which allows imaging the standing waves. The lateral periodicity corresponds as expected
to half the wavelength in the wave guide ($\lambda/2n\approx500\mathrm{nm}$).
Note that in this vertical coupling configuration we observe scattering due to the nanostring which is however unequally distributed among the two modes since they can exhibit either a maximum (leading to non-negligible scattering) or a node (leading to negligible scattering) at the nanomechanical oscillator's position.
At the location in Fig. S \ref{SIf:4}c indicated by the red arrow (corresponding to the transmission plotted in panel b) the system exhibits a quadratic optical frequency shift with respect to the transverse motion $y$ of the string, i.e. a nonzero $\frac{\partial^{2}}{\partial y^{2}}
\omega_\mathrm{0}(y)=g^{(2)}$, while $\frac{\partial}{\partial y}\omega_\mathrm{0}(y)=0$. Hence, the
interaction Hamilitonian is quadratic \cite{SIThompson08,SIMiaoA09}, i.e. $\widehat
{H}_{\mathrm{int}}=\left(\hbar g^{(2)}\,y_\mathrm{zp}^{2}/2\right)\,\hat{n}_m\hat{a}^{\dagger}\hat{a}$, where
$y_\mathrm{zp}$ is the mechanical zero point motion and $\hat{n}_m$ the number operator of the
mechanical oscillator. The latter is of interest in order to achieve an
experimental setting in which a QND measurement of mechanical
quanta \cite{SIThompson08,SIMiaoA09} becomes in principle feasible.
\clearpage


\end{document}